# *Ab initio* superionic-liquid phase diagram of $Fe_{1-x}O_x$ under Earth's inner core conditions

Zepeng Wu[1], Chen Gao[1], Feng Zhang[2], Shunqing Wu[1], Kai-Ming Ho[2], Renata M. Wentzcovitch[3-6], Yang Sun[1]

[1]*Department of Physics, Xiamen University, Xiamen 361005, China*
[2]*Department of Physics, Iowa State University, Ames, IA 50011, USA*
[3]*Department of Applied Physics and Applied Mathematics, Columbia University, New York, NY 10027, USA*
[4]*Department of Earth and Environmental Sciences, Columbia University, New York, NY 10027, USA*
[5]*Lamont–Doherty Earth Observatory, Columbia University, Palisades, NY 10964, USA*
[6]*Data Science Institute, Columbia University, New York, NY 10027, USA*

The superionic state is a phase of matter in which liquid-like ionic mobility coexists with a solid crystalline lattice. Recently identified in Earth's inner core (IC), this state has attracted considerable attention for its unique kinetic behavior and geophysical implications. However, the *ab initio* phase diagram describing the equilibrium between the superionic phase and the liquid solution under core conditions remains largely unexplored. Here, we present a thermodynamic approach to compute the Gibbs free energy and construct the *ab initio* superionic–liquid phase diagram for the $Fe_{1-x}O_x$ system under IC conditions. We find that oxygen forms superionic states in both *hcp* and *bcc* Fe phases, with a pronounced influence on cooperative diffusion of iron in the *bcc* lattice. The stability fields of these superionic phases are sensitive to oxygen stoichiometry. The presence of superionic states leads to a higher oxygen concentration in the IC than previously estimated. Our work establishes a framework for investigating superionic–liquid equilibria under extreme conditions.

Superionicity is a unique state where materials exhibit liquid-like mobility within a crystal lattice, drawing great interest in various scientific and industrial fields. Under ambient conditions, this property is crucial for solid electrolytes vital for next-generation all-solid-state batteries [1,2]. At extremely high pressures and temperatures, solid phases can transform into superionic states [3–7], as is believed to occur in minerals of Earth's deep mantle [8–10], as well as in ice and ammonia within the mantles of Uranus and Neptune [11–17]. Recent simulations reveal that light elements like oxygen, hydrogen, and carbon can become superionic in hexagonal close-packed (*hcp*) iron under Earth's inner core (IC) conditions [18]. The kinetic behavior of the superionic state is proposed to cause the anisotropic seismic characteristics of the IC [19]. However, the stability field of the superionic state in the IC remains unclear. The phase competition between *hcp* and body-centered cubic (*bcc*) iron under IC conditions has long been debated [20,21]. While recent studies suggest *hcp* as the stable phase [22,23], it is uncertain if superionic states can emerge in *bcc* Fe-light element alloys and affect their stability relative to superionic *hcp* alloys under IC conditions. As the compositions of light elements were determined based on the solid solution models of the *hcp* phase [24–26], superionic solutions could alter our understanding of light element partitioning between the solid IC and the liquid outer core. Thus, determining the thermodynamic stability of superionic phases, especially their equilibrium with liquid solutions at the inner core boundary (ICB), is crucial not only for fundamental physics in the novel state but also for constraining the core's structure and chemical composition, which are vital for understanding the Earth's deep interior [20,27].

Despite its importance, exploring the stability of the superionic phase in the IC is challenging. Experimental observation of superionicity in Fe alloys is lacking due to the difficulty of detecting this state under IC conditions. Theoretical studies of phase competition among liquid, superionic, and solid phases are scarce, as calculating the free energy of the superionic state is highly non-trivial [28]. In the case of superionic ice, several attempts have been proposed to compute its free energy. A typical method is based on thermodynamic integration (TI), which provides the difference in free energy between the superionic phase and a reference model for which absolute free energy is known as *a priori*. Although TI is accurate, finding a suitable reference for the superionic phase is difficult. Wilson et al. proposed using noninteracting harmonic oscillators and an ideal gas as a superionic reference [29]. However, this model suffers from the problem of particle overlap due to the lack of interactions between solid-like and liquid-like particles [30]. Cheng et al. used machine learning interatomic potentials to simulate the superionic and liquid coexistence for stoichiometric ice phases, providing melting curves for superionic ice [14]. While simulations of coexisting phases are sufficient to establish phase equilibria for stoichiometric systems such as $H_2O$, as we will demonstrate later, direct calculations of the absolute free energy for the liquid and superionic phases are necessary to obtain a complete phase diagram for the non-stoichiometric $Fe_{1-x}O_x$ system. Moreover, classical simulations strongly depend on the accuracy of the interatomic potentials, even trained by machine learning techniques [31]. An approach that eliminates dependence on interatomic potential accuracy and enables the determination of the superionic phase's free energy at the *ab initio* level remains desirable. In principle, the interatomic potential can serve as a reference model for computing the *ab initio* Gibbs free energy of the superionic state in TI. A similar method was recently developed to determine the melting temperatures and relative free energies of pure Fe phases under IC conditions [22], showing that classical simulations provide a suitable reference for *ab initio* calculations using classical-to-*ab initio* thermodynamic integration (CATI) and free-energy perturbation (FEP) methods [22,32]. While this method is relatively straightforward for systems with constant stoichiometry like $H_2O$ or Fe, it becomes complex for non-stoichiometric solutions where the chemical potential of solute atoms in liquid and superionic phases is critical.

In this work, we develop a scheme to calculate the *ab initio* superionic-liquid phase diagram for the non-stoichiometric $Fe_{1-x}O_x$ system under conditions close to Earth's ICB, considering the superionic state in both *hcp* and *bcc* phases. We first construct a Fe-O interatomic potential to simulate the coexistence of superionic and liquid phases in $Fe_{1-x}O_x$ systems using large-scale, long-timescale molecular dynamics (MD) simulations, which provide solidus and liquidus curves. Based on thermodynamic relations, we show that the *absolute* Gibbs free energy of the superionic phase can be computed without any assumption. Using CATI and FEP methods, we obtain the *ab initio* Gibbs free energy of both liquid and superionic phases from the interatomic-potential reference. By introducing the realistic physical reference for the superionic state, we eliminate the particle-overlap problem in previous thermodynamic integration calculations [29,30]. The framework (Supplementary Fig. S1 [33]) allows us to construct a superionic-liquid phase diagram with *ab initio* accuracy. We elucidate phase competition between superionic *hcp* and *bcc* structures and assess the impact of the superionic state on O partitioning between inner and outer cores.

*Effect of superionic oxygen on Fe's cooperative motion*—The key feature of the superionic $Fe_{1-x}O_x$ alloy is the O diffusion in the crystalline lattice. Based on the AIMD simulation, O exhibits superionic behavior in both *hcp* and *bcc* Fe under ICB conditions. It shows a similar mean squared displacement (MSD) in *hcp*, *bcc,* and liquid Fe phases at 323 GPa and 5500 K, conditions close to those at ICB (Supplementary Fig. S4 [33]). In the *bcc* lattice, Fe atoms exhibit cooperative diffusion motion under IC conditions, where atoms move in a string-like manner among crystalline sites [21,34]. We quantify the Fe's motion using the van Hove self-correlation function $G_s(r,\Delta t) = \frac{1}{N}\langle \sum_{i=1}^{N} \delta\left(\vec{r} + \vec{r}_i(0) - \vec{r}_i(\Delta t)\right)\rangle$, where $\Delta t$ is chosen to be 6 *ps*, which can well distinguish the vibrational motion from the cooperative diffusion motion. As shown in Fig. 1(a), the second peak of Fe's van Hove self-correlation function at 2.1 Å systematically increases with rising oxygen concentration. This peak corresponds to Fe's cooperative motion with its nearest neighboring Fe atom. Moreover, a third peak at 3.8 Å emerges when $x_O$ exceeds 1.19%, suggesting multiple cooperative motions within the time period. Thus, with higher oxygen concentrations, Fe's diffusion motion becomes more pronounced. Figure 1(b) shows that Fe atoms exhibiting cooperative motion have more oxygen neighbors than the average distribution. The trajectory in Fig. 1(b) also provides a clear visualization of this phenomenon. Thus, superionic oxygen enhances Fe's cooperative diffusion motion in the *bcc* phase under IC conditions. This can be attributed to the fact that O enhances the tendency of Fe atoms to deviate from their crystalline sites, as indicated by the phonon instability analysis in the bcc lattice (Supplementary Note 1 [33]).

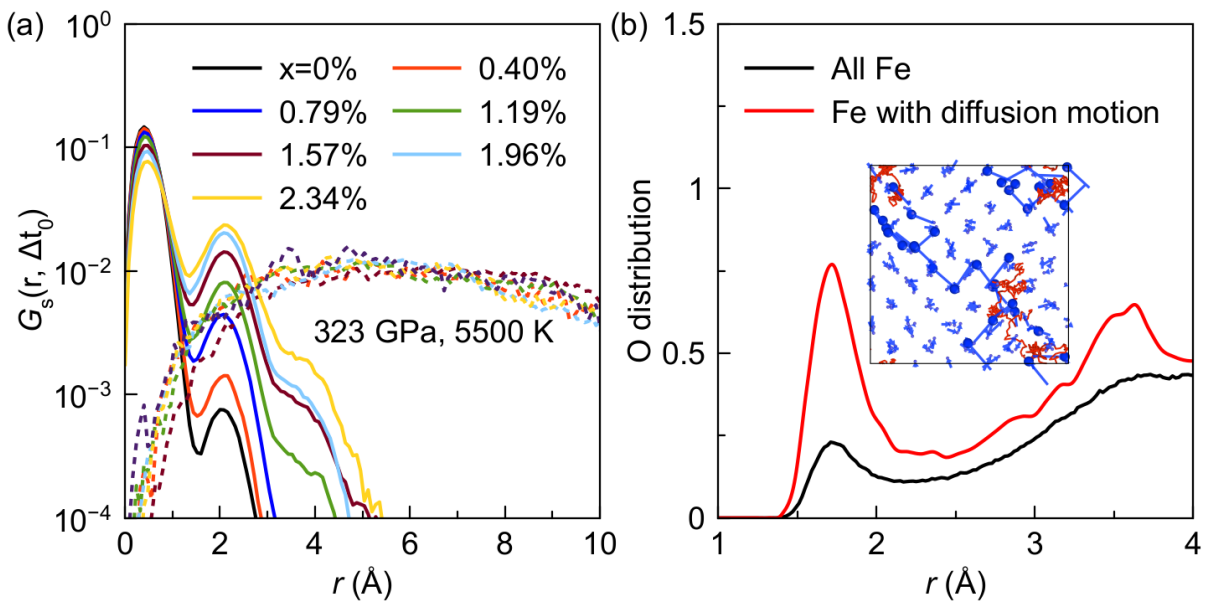

FIG. 1. The effect of superionic O on the Fe's cooperative diffusion in *bcc*. (a) The van Hove self-correlation function for $Fe_{1-x}O_x$ *bcc* phases at 323 GPa and 5500 K. (b) The radial distribution function of O surrounding Fe atoms. Red line is for the Fe exhibiting cooperative diffusion motion. The data are collected in 1.5 *ps* prior to the completion of Fe's cooperative diffusion motion. The insert shows the trajectory of Fe's cooperative diffusion motion (blue) and O's superionic motions (red). The Fe atoms exhibiting only vibrational motions are reduced in size for clarity.

*Superionic-liquid coexistence*—Because *ab initio* MD simulation is highly limited by the time and length scales, we employ classical MD simulations to study the stability of superionic phases coexisting with liquid. We developed a Fe-O interatomic potential using the embedded-atom method (EAM) [35], which can simulate the superionic state in both *hcp* and *bcc* lattices without any instability under ICB conditions. The MSD from classical MD align qualitatively well with *ab initio* data (Supplementary Note 2 [33]). To simulate the superionic *hcp*-liquid coexistence, we constructe a two-phase model with an *hcp*-liquid interface using 12,288 Fe atoms and randomly distributed O atoms for various O compositions. After MD simulations, O composition decreases in the *hcp* phase as shown in Fig. 2(a). The system reaches equilibrium at around 100 *ps*, as indicated by the energy change in Fig. 2(b). With a long simulation time of 2 *ns*, O atoms diffused throughout the simulation cell, providing sufficient data to compute their equilibrium composition in the *hcp* and liquid phases in Fig. 2(c). The O composition in the superionic phase ($x_C^{SI}$) is significantly lower than that in the liquid phase ($x_C^{L}$). Tests with larger system sizes suggest the current size is sufficient to obtain converged results (Supplementary Fig. S8 [33]). Similar simulations across temperatures from 4600-6000 K at the same pressure yielded temperature-dependent $x_C^{SI}$ and $x_C^{L}$, representing the superionic *hcp* solidus and liquidus lines in Fig. 2(d).

Based on the thermodynamic relations, when $Fe_{1-x}O_x$ superionic solution coexists in equilibrium with the liquid solution, it satisfies the equilibrium condition,

$$G_C^{SI}(x_{C0}^{SI}) = G_C^L(x_{C0}^L) - (x_{C0}^L - x_{C0}^{SI})\frac{\partial G_C^L(x)}{\partial x}\Big|_{x=x_{C0}^L} \quad (1)$$

where $G_C^L(x_{C0}^L)$ and $G_C^{SI}(x_{C0}^{SI})$ are absolute Gibbs free energies of liquid and superionic phases with O compositions of $x_{C0}^L$ and $x_{C0}^{SI}$, respectively. The detailed derivations of Eqn. (1) are presented in End Matter. Since the MD simulation in Fig. 2(d) provides $x_{C0}^L$ and $x_{C0}^{SI}$ at various temperatures $T_0$, Eqn. (1) can be employed to compute the Gibbs free energy of the superionic phase $G_C^{SI}(x_{C0}^{SI}, T_0)$, provided the liquid's Gibbs free energy $G_C^L(x, T_0)$ is known. The nonequilibrium TI method was employed to compute the Helmholtz free energy for the liquid solutions (Supplementary Note 3 [33]). A series of free energy calculations for a liquid solution are performed across the O composition range of 0-20 at.%, with the compositions spaced equally, at various temperatures in Fig. 3(a). We find these liquid's free energy data can be well described by the regular solution model using the Redlich-Kister (RK) expansion [36] as

$$G(x,T_0) = G^{Fe}(T_0) + ax + x(1-x)\sum_{k=0}^{n_k} L_k(2x-1)^k + k_B T_0[x\ln x + (1-x)\ln(1-x)]\,, \quad (2)$$

where $G^{Fe}(T_0)$ is the Gibbs free energy of pure Fe. $a$ and $L_k$ are the fitting parameters. It only requires two RK terms ($k = 0$ and 1) to fit the liquid's free energy data, achieving fitting errors of less than 0.1 meV/atom at all temperatures studied.

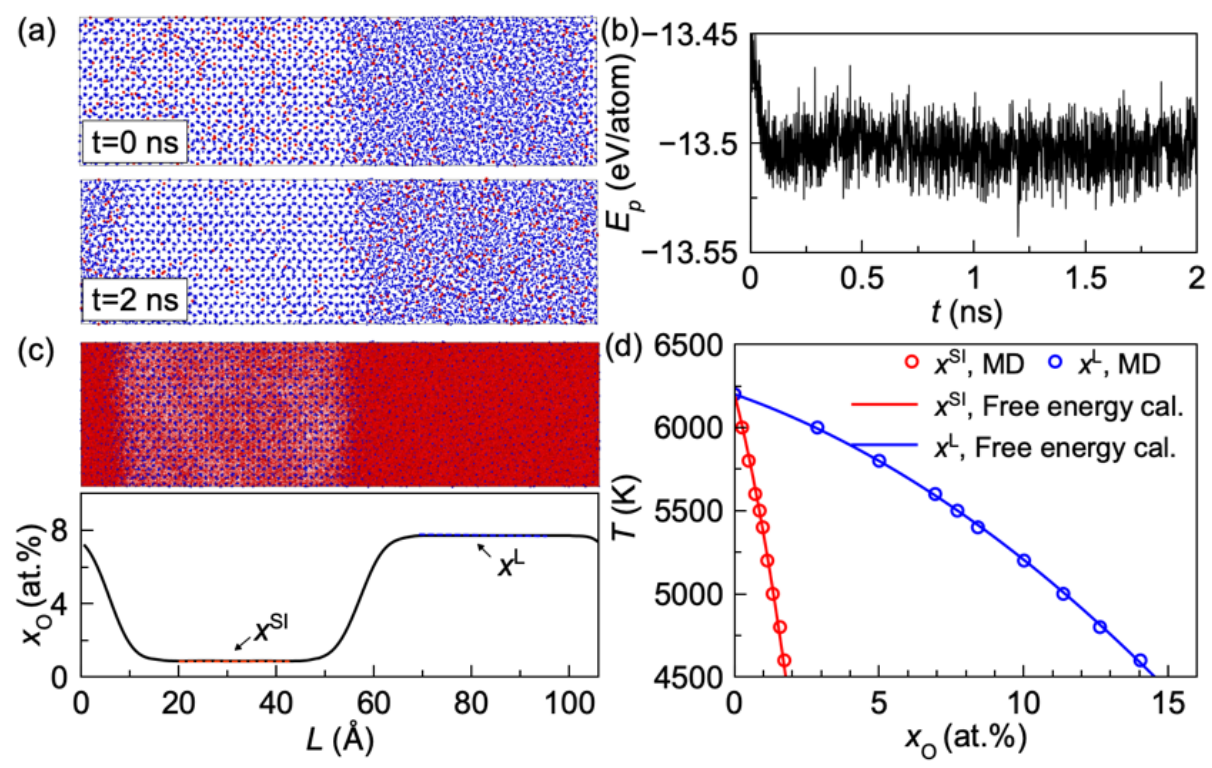

FIG. 2. Superionic-liquid coexistence simulations of $Fe_{1-x}O_x$ solution. (a) Initial and final configurations the coexistence simulation at 323 GPa and 5500 K. The blue and red dots represent Fe and O atoms, respectively. (b) Potential energy as a function of time in the simulation. (c) O trajectory in last 1 *ns*. The lower panel shows the averaged O composition along the direction perpendicular to the interface. (d) The phase diagram by classical MD simulation at 323 GPa. The circles are from direct superionic-liquid coexistence simulations. The lines are from free energy calculations.

Based on Eqn. (1) and $G_C^L(x, T_0)$, we can compute $G_C^{SI}(x_{C0}^{SI}, T_0)$ for each $(x_{C0}^{SI}, x_{C0}^L, T_0)$ combination obtained from superionic-liquid coexistence simulations in Fig. 2(d). This results in sparse Gibbs free energy data for the superionic *hcp* state at a few temperatures, marked as solid circles in Fig. 3(b). We then extend these data to a broader temperature range using the Gibbs-Helmholtz equation,

$$G(x_0,T) = \frac{T}{T_0}G(x_0,T_0) - T\int_{T_0}^{T}\frac{H(x_0,T)}{T^2}dT, \quad (3)$$

where $H(x_0, T)$ is the temperature-dependent enthalpy with a specific O composition of $x_0$ obtained from MD simulations. Temperature-dependent $G_C^{SI}(x_{C0}^{SI}, T)$ are computed with Eqn. (3) for different $x_{C0}^{SI}$ and plotted as a function of compositions in Fig. 3(b). The free energy of superionic phase can also be well fitted by Eqn. (2) (Supplementary Note 4 [33]) using only one RK term ($k = 0$) to achieve fitting errors of less than 0.2 meV/atom (Supplementary Fig. S9 [33]).

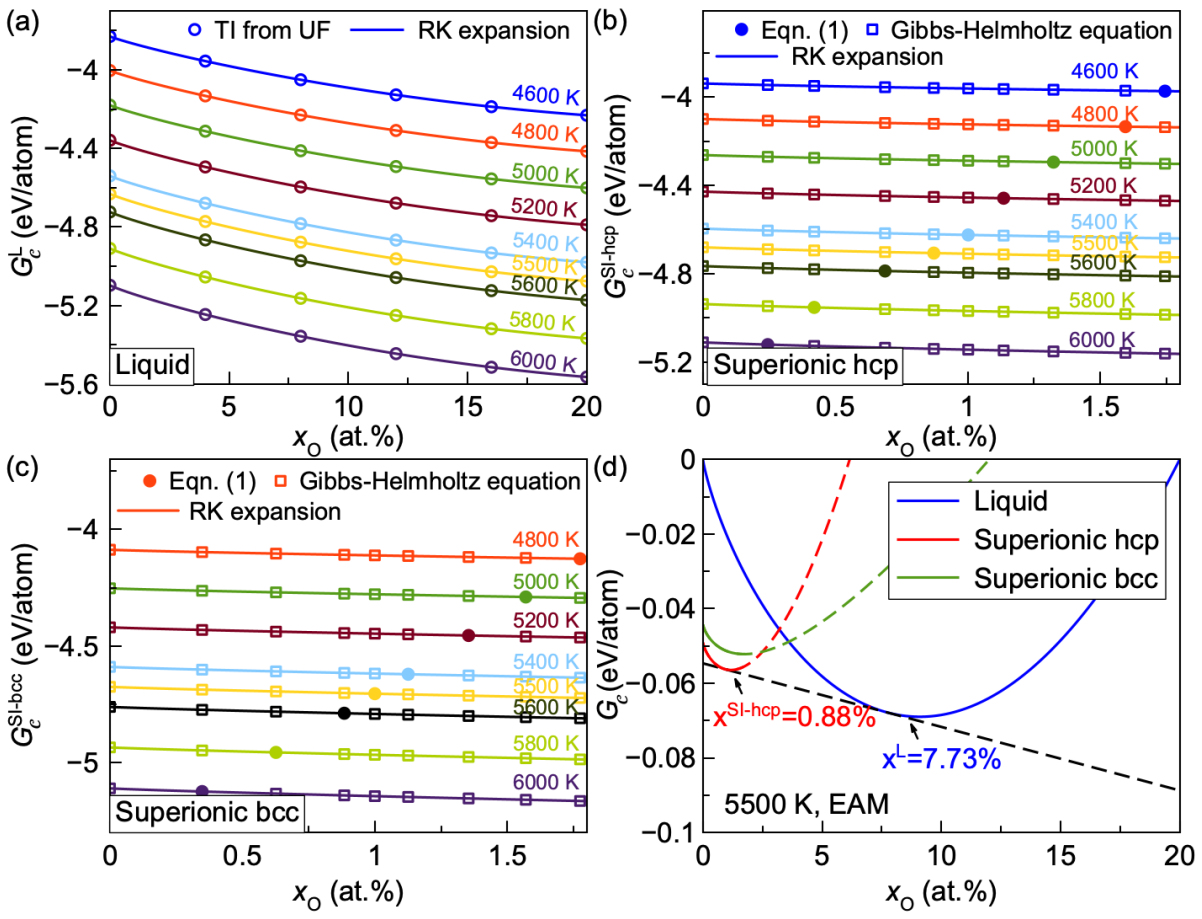

FIG. 3. (a) The composition-dependent Gibbs free energy of liquid solutions with EAM potential at 323 GPa. (b) The composition-dependent Gibbs free energy of superionic *hcp* solutions. The solid circles are computed based on the superionic-liquid equilibrium condition in Eqn. (1). The open squares are computed based on the Gibbs-Helmholtz equation and the values of the solid circles. The solid lines represent the fitting with the RK expansion. (c) The composition-dependent Gibbs free energy of superionic *bcc* solutions. (d) The relative Gibbs free energy for liquid and superionic solutions at 323 GPa and 5500 K from EAM potential. The solid (dashed) line indicates the interpolated (extrapolated) results using RK expansion. The dashed line is the common tangent between liquid and superionic *hcp* data, defining solidus and liquidus compositions. The data are referenced to 0% and 20% liquid free energy for better visualization.

With the absolute Gibbs free energy for both liquid and superionic solutions across various compositions and temperatures, the common tangent line can be computed to get the solidus and liquidus curves. Figure 3(d) shows the common tangent lines computed between $G_C^{SI-hcp}(x)$ and $G_C^L(x)$ curves at 5500 K, which resulted in intersections at $x_{C0}^{SI-hcp} = 0.88\%$ and $x_{C0}^L = 7.73\%$. These values are consistent with the equilibrium compositions of $x_{C0}^{SI-hcp} = 0.87\%$ and $x_{C0}^L = 7.70\%$ obtained from superionic *hcp*-liquid coexistence simulations under the same pressure and temperature conditions shown in Fig. 2(c). More free energy data and their common tangent lines at other temperatures are shown in Supplementary Fig. S14 [33]. The superionic *hcp* solidus and liquidus curves computed from the free energy calculations are compared with those from coexistence simulations in Fig. 2(d). Both methods result in a consistent superionic-liquid phase diagram, validating each other.

Using Eqns. (1)-(3), we repeated the calculations for the superionic *bcc* phase. The temperature-dependent free energies of superionic *bcc* are shown in Fig. 3(c). The solidus and liquidus curves of superionic *bcc* are shown in Fig. S15 [33]. Figure 3(d) and Supplementary Fig. S14 [33] compare the relative free energy among liquid, superionic *bcc*, and superionic *hcp* phases at 5500 K. These data suggest that the superionic *bcc* is metastable compared to the superionic *hcp* phase when the O composition is small. When the O composition is greater than ~3 at.%, the superionic *bcc* becomes more stable than superionic *hcp* in $Fe_{1-x}O_x$.

*Ab initio Gibbs free energy of superionic phases*—We have obtained the absolute Gibbs free energy, $G_{\mathcal{C}}$, of the liquid and superionic solutions for the EAM system. Using the classical EAM system as the reference state, we can perform CATI [22,37] to compute the *ab initio* Gibbs free energy $G_{\mathcal{A}}$ by

$$G_{\mathcal{A}} = G_{\mathcal{C}} + f_{PV} + f_{TI}, \tag{4}$$

where $f_{TI}$ is the Helmholtz free energy difference and $f_{PV}$ is the PV contribution. Please refer to End Matter for detailed derivations in Eqn. (A8-A13). The accuracy of the CATI scheme is tested to be independent of the specific EAM potential (see Supplementary Note 5 [33]), as long as the reference potential can be smoothly connected to the *ab initio* state without phase transitions. A large amount of AIMD simulations were performed to compute the equilibrium volumes for liquid, superionic *hcp,* and *bcc* solutions at 5500-6000 K (Supplementary Fig. S16 [33]), which provides the $f_{PV}$ term in Eqn. (4). To obtain $f_{TI}$, a series of TI simulations from the classical system to the *ab initio* system were performed for liquid, superionic *hcp,* and superionic *bcc* solutions at 5500 K, as shown in Fig. 4(a). The TI path from the $\mathcal{C}$ to the $\mathcal{A}$ system is smooth, significantly reducing numerical errors in TI and enabling accurate determination of the free energy difference. With Eqn. (4), $G_{\mathcal{A}}$ can be calculated for liquid, superionic *hcp,* and *bcc* at different O compositions (Supplementary Fig. S17 [33]). We also perform the free energy perturbation to include contributions of $3s^2 3p^6$ electrons, which have been shown to be necessary for accurately calculating the free energy of Fe under IC conditions [22,38] (Supplementary Fig. S18 [33]).

Figure 4(b) shows the relative *ab initio* Gibbs free energy for liquid and superionic phases at 5500 K and 323 GPa. The regular solution model with the RK expansion in Eqn. (2) can also well describe these free energies, providing fitting errors of less than 0.5 meV/atom (see Supplementary Fig. S9 [33]). Figure 4(b) shows the *ab initio* free energy of the superionic *bcc* phase is higher than that of the superionic *hcp* when the O composition is small. This is consistent with the fact that pure Fe prefers the *hcp* phase under IC conditions [22,23]. As the O composition increases, the free energy of the superionic *hcp* quickly increases. When the O composition is higher than 3 at.%, the superionic *bcc* phase becomes more stable than the superionic *hcp* phase. This can be related to the enhancement of Fe cooperative motion due to superionic O (Fig. 1), which increases the entropy of the *bcc* phase (Supplementary Note 8 [33]). Thus, the O composition in $Fe_{1-x}O_x$ changes the relative stability between the superionic *bcc* and *hcp* phases.

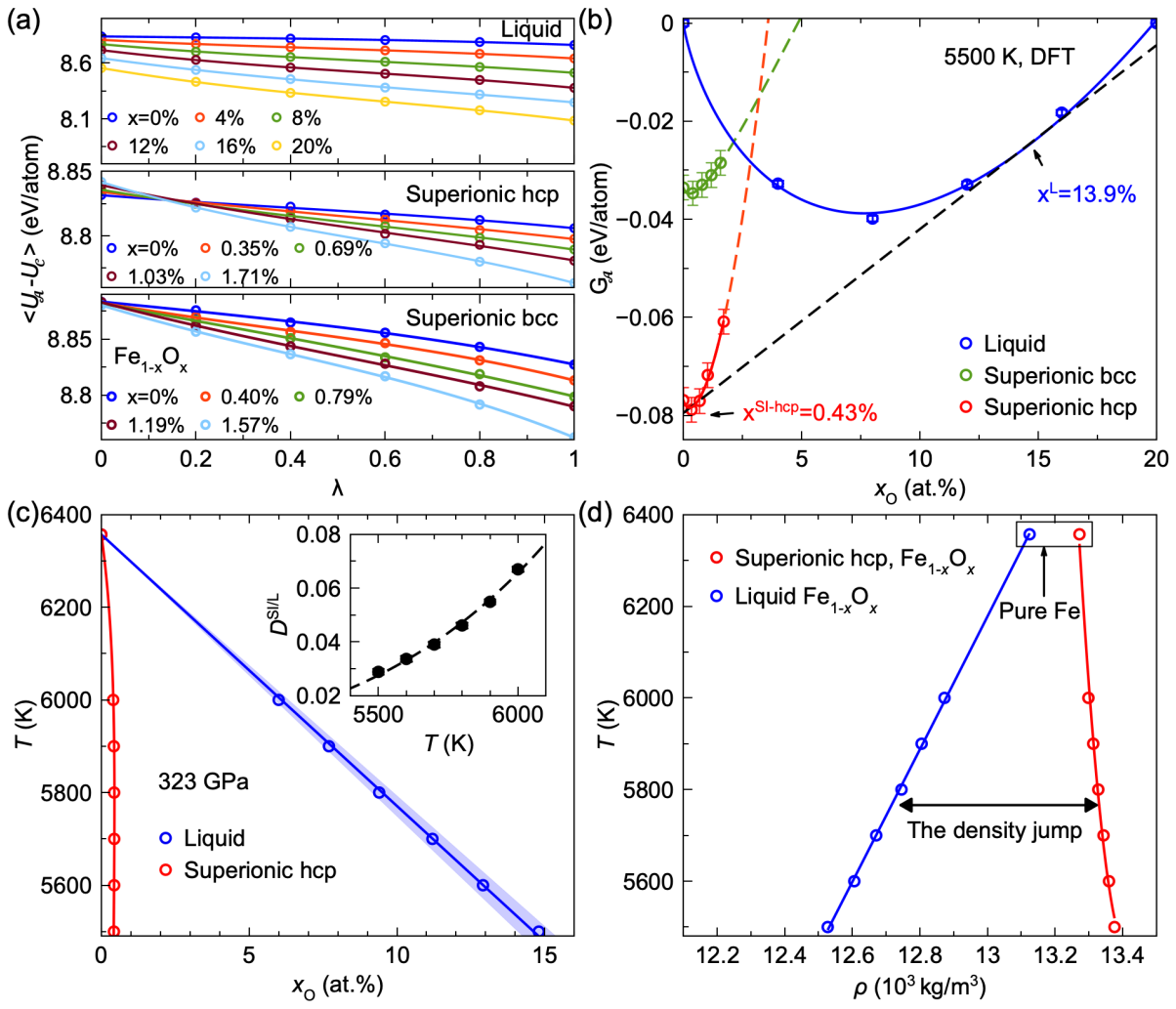

FIG. 4. *Ab initio* phase diagram of superionic and liquid $Fe_{1-x}O_x$ solutions. (a) The energy differences in the TI calculation from the classical to the *ab initio* system. The lines are the third-order polynomial fitting. (b) The relative *ab initio* Gibbs free energy for liquid, superionic *hcp*, and superionic *bcc* solutions. The solid and dashed lines indicate the interpolated and extrapolated results using RK expansion, respectively. The black dashed line is the common tangent line. (c) The superionic-liquid phase diagram of the $Fe_{1-x}O_x$ superionic-liquid system at the *ab initio* level at 323 GPa. The inset shows the temperature dependent partition coefficient, $D^{SI/L} = x_{\mathcal{A}}^{SL}/x_{\mathcal{A}}^{L}$. The shaded areas represent the uncertainty ranges of $x^L$ and $x^{SI}$. (d) Density of superionic *hcp* and liquid $Fe_{1-x}O_x$ under the equilibrium composition at different temperatures.

The common tangent line between liquid and superionic phases suggests that the superionic solution can only coexist with the liquid solution at small O compositions. Based on the $G_{\mathcal{A}}^{SI-hcp}(x)$ and $G_{\mathcal{A}}^{L}(x)$ curves, the common tangent line reveals *ab initio* solidus and liquidus points at 5500 K are $x_{\mathcal{A}}^{SI-hcp} = 0.43\%$ and $x_{\mathcal{A}}^{L} = 13.9\%$, respectively. Superionic *hcp* is more stable than superionic *bcc* at such a small O composition. We employ Eqn. (3) to extend the free energy data to other temperatures for all three phases (Supplementary Fig. S19 [33]). The superionic *hcp* solidus and liquidus curves are computed using the common tangent approach. These data provide the *ab initio* phase diagram of superionic *hcp* and liquid $Fe_{1-x}O_x$ at 323 GPa in Fig. 4(c), with uncertainties analyzed in Supplementary Note 6 [33]. The superionic solidus line shows small temperature dependences, while the liquidus line depends on the temperature more strongly. It results in a partition coefficient strongly dependent on temperatures.

*Oxygen concentration at ICB*—The phase diagram in Fig. 4(c) indicates that the equilibrium O compositions in the solid IC and liquid outer core are highly correlated with the temperature at ICB. It provides a stronger constraint on the core's composition and temperature than the one from the partition coefficient data alone. Based on the phase diagram, the densities of $Fe_{1-x}O_x$ superionic and liquid solutions can be computed under the equilibrium conditions at different temperatures, as shown in Fig. 4(d). In previous work, substitutional O was introduced to account for the large density difference between the liquid and solid cores, i.e., the "density jump", resulting in estimated O concentrations of 0.2±0.1 at.% in the substitutional solid and 8.0±2.5 at.% in the liquid solutions [39]. To assess the effect of superionic O, we follow the assumption in [39] by matching the density jump in PREM [40] using only O, but based on the calculated equilibrium concentrations and densities of superionic *hcp* and liquid $Fe_{1-x}O_x$ solutions. It simultaneously constrains the O concentrations to 0.432±0.003 at.% (0.124±0.001 wt.%) in the superionic solid and 9.82±1.45 at.% (3.02±0.42 wt.%) in the liquid, along with a temperature of 5770±86 K at the ICB. Therefore, compared to the previous substitutional solid solution, the superionic phase doubles O's solubility in *hcp* Fe under IC conditions. This extends the previously estimated range of O's concentration in the IC (0–0.1 wt.%), which are based on constraints from cosmochemistry, geochemistry, and mineral physics [24]. Our results add a missing piece by revealing the influence of the superionic state on these estimates. A different oxygen content in the IC can alter the energy release and thermal evolution of the core, potentially influencing core growth dynamics and the efficiency of compositional convection in the outer core, both essential for sustaining the geodynamo.

Our kinetic and thermodynamic data clarify the effect of superionic O on both the *hcp* and *bcc* phases. Since the *bcc* phase becomes more stable at O concentrations above 3 at.%, which likely fall within the estimated O content of the liquid core [24,41], this suggests that IC crystallization may begin with superionic *bcc* $Fe_{1-x}O_x$, especially considering the *bcc* phase exhibits a much higher nucleation rate than hcp [42]. Moreover, recent studies have shown that elements such as Ni and Si can further stabilize the *bcc* phase [43–46]. Based on these findings and our free energy data, mutual interactions among Ni, Si, and other light elements are expected to further influence both superionic O concentration and the stable crystalline phase in the IC. Therefore, *ab initio* Gibbs free energy calculations and phase diagrams of multicomponent Fe alloys—including the superionic state and various crystalline phases—are essential, and the present work provides a solid framework for such investigations.

In summary, we demonstrate that O can enter a superionic state in both *hcp* and *bcc* Fe under Earth's IC conditions. In particular, the superionic O enhances the cooperative diffusion of Fe atoms in the *bcc* lattice, increasing the entropy of the *bcc* phase and influencing the phase competition between *hcp* and *bcc* Fe. To accurately capture the thermodynamics of these complex phases, we develop an effective approach that combines classical coexistence simulations with *ab initio* free energy calculations to construct phase diagrams for non-stoichiometric superionic states. Applied to the $Fe_{1-x}O_x$ under IC conditions, our method quantitatively determines the Gibbs free energies of superionic *bcc*, superionic *hcp*, and liquid phases. The combination between the phase diagram and the density jump data simultaneously constrains the O composition and temperature at the ICB. Our results show that the presence of the superionic state nearly doubles the O content in *hcp* Fe compared to previous solid solution models. Moreover, superionic O emerges as an additional stabilizing factor that lowers the free energy of *bcc* Fe at elevated O concentrations. The resulting Gibbs free energy data and superionic-liquid phase diagram provide new constraints on the composition and temperature of the IC, underscoring the critical role of superionicity in shaping Earth's core structure and evolution. The framework developed here establishes a foundation for investigating superionic–liquid equilibria in multicomponent Fe alloys under planetary interior conditions.

*Acknowledgments*—Work at Xiamen University was supported by the National Natural Science Foundation of China (Grants T2422016, 42374108 and 12374015). The work at Columbia University was supported by the Gordon and Betty Moore Foundation Award GBMF12801 (doi.org/10.37807/GBMF12801). Shaorong Fang and Tianfu Wu from the Information and Network Center of Xiamen University are acknowledged for their help with Graphics Processing Unit computing. Some *ab initio* simulations were performed on Bridges-2 system at PSC, the Anvil system at Purdue University, the Expanse system at SDSC, and the Delta system at NCSA through allocation DMR180081 from the Advanced Cyberinfrastructure Coordination Ecosystem: Services & Support (ACCESS) program, which is supported by NSF Grants No. 2138259, No. 2138286, No. 2138307, No. 2137603, and No. 2138296. The supercomputing time were also supported by the Opening Project of the Joint Laboratory for Planetary Science and Supercomputing, Research Center for Planetary Science, and the National Supercomputing Center in Chengdu (Grants No. CSYYGS-QT-2024-15).

## End Matter

*Free energy relation in the superionic-liquid equilibrium*—Let's consider the thermodynamic equilibrium in a superionic-liquid coexistence system. We use $G^L(x^L,T,P)$ and $G^{SI}(x^{SI},T,P)$ to represent the absolute Gibbs free energy of the liquid and superionic phases, respectively. $x^L$ and $x^{SI}$ represent the O content in the liquid and superionic $Fe_{1-x}O_x$, respectively. The Gibbs free energy of the liquid and superionic solutions at temperature $T$ and pressure $P$ can be expressed as

$$G^L(x^L,T,P) = x^L\bar{G}_O^L(x^L,T,P) + (1-x^L)\bar{G}_{Fe}^L(x^L,T,P), \quad \text{(A1a)}$$

$$G^{SI}(x^{SI},T,P) = x^{SI}\bar{G}_O^{SI}(x^{SI},T,P) + (1-x^{SI})\bar{G}_{Fe}^{SI}(x^{SI},T,P), \quad \text{(A1b)}$$

where $\bar{G}_O^L(x^L,T,P)$ and $\bar{G}_{Fe}^L(x^L,T,P)$ are the partial molar Gibbs free energy of oxygen and iron in the liquid solutions, respectively. $\bar{G}_O^{SI}(x^{SI},T,P)$ and $\bar{G}_{Fe}^{SI}(x^{SI},T,P)$ are the partial molar Gibbs free energy of oxygen and iron in the superionic solutions, respectively. Taking the derivative of both sides of Eqn. (A1a) and (A1b) with respect to $x^L$ and $x^{SI}$ and the Gibbs-Duhem relation, we get

$$\frac{\partial G^L(x,T,P)}{\partial x}\Big|_{x=x^L} = \bar{G}_O^L(x^L,T,P) - \bar{G}_{Fe}^L(x^L,T,P), \quad \text{(A2a)}$$

$$\frac{\partial G^{SI}(x,T,P)}{\partial x}\Big|_{x=x^{SI}} = \bar{G}_O^{SI}(x^{SI},T,P) - \bar{G}_{Fe}^{SI}(x^{SI},T,P). \quad \text{(A2b)}$$

By combining Eqn. (A1) and (A2) and eliminating $\bar{G}_{Fe}^L(x^L,T,P)$ and $\bar{G}_{Fe}^{SI}(x^{SI},T,P)$, we can obtain the partial molar Gibbs free energy of oxygen in both liquid and superionic $Fe_{1-x}O_x$ as follows

$$\bar{G}_O^L(x^L,T,P) = G^L(x^L,T,P) + (1-x^L)\frac{\partial G^L(x,T,P)}{\partial x}\Big|_{x=x^L}, \quad \text{(A3a)}$$

$$\bar{G}_O^{SI}(x^{SI},T,P) = G^{SI}(x^{SI},T,P) + (1-x^{SI})\frac{\partial G^{SI}(x,T,P)}{\partial x}\Big|_{x=x^{SI}}. \quad \text{(A3b)}$$

By combining Eqn. (A2) and (A3), the partial molar Gibbs free energy of iron in liquid and superionic solutions are as follows

$$\bar{G}_{Fe}^L(x^L,T,P) = G^L(x^L,T,P) - x^L\frac{\partial G^L(x,T,P)}{\partial x}\Big|_{x=x^L}, \quad \text{(A4a)}$$

$$\bar{G}_{Fe}^{SI}(x^{SI},T,P) = G^{SI}(x^{SI},T,P) - x^{SI}\frac{\partial G^{SI}(x,T,P)}{\partial x}\Big|_{x=x^{SI}}. \quad \text{(A4b)}$$

When the liquid and superionic solutions reach equilibrium at the temperature $T_0$ and pressure $P_0$, the partial molar Gibbs free energy of both iron and oxygen in both phases are equal [47]. Therefore, they satisfy

$$\bar{G}_O^L(x_0^L,T_0,P_0) = \bar{G}_O^{SI}(x_0^{SI},T_0,P_0), \quad \text{(A5a)}$$

$$\bar{G}_{Fe}^L(x_0^L,T_0,P_0) = \bar{G}_{Fe}^{SI}(x_0^{SI},T_0,P_0), \quad \text{(A5b)}$$

where $x_0^L$ and $x_0^{SI}$ are the oxygen contents in liquid and superionic solutions when the coexistence system reaches equilibrium at the fixed temperature $T_0$ and pressure $P_0$. So that we have,

$$\frac{\partial G^L(x,T_0,P_0)}{\partial x}\Big|_{x=x_0^L} = \frac{\partial G^{SI}(x,T_0,P_0)}{\partial x}\Big|_{x=x_0^{SI}}, \quad \text{(A6a)}$$

$$G^L(x_0^L,T_0,P_0) - x_0^L\frac{\partial G^L(x,T_0,P_0)}{\partial x}\Big|_{x=x_0^L} = G^{SI}(x_0^{SI},T_0,P_0) - x_0^{SI}\frac{\partial G^{SI}(x,T_0,P_0)}{\partial x}\Big|_{x=x_0^{SI}}. \quad \text{(A6b)}$$

By Eqn. (A6), the absolute Gibbs free energy of the superionic with an oxygen molar fraction $x_0^{SI}$ is as follows

$$G^{SI}(x_0^{SI},T_0,P_0) = G^L(x_0^L,T_0,P_0) - (x_0^L - x_0^{SI})\frac{\partial G^L(x)}{\partial x}\Big|_{x=x_0^L}. \quad \text{(A7)}$$

Eqn. (A7) indicates that if the oxygen concentrations in the superionic-liquid equilibrium and the liquid free energy are known, the Gibbs free energy of the superionic phase can be directly calculated. It is straightforward to obtain these quantities using large-scale MD simulations with interatomic potentials.

*Thermodynamic integration from classical to ab initio systems*—When the Gibbs free energy $G_{\mathcal{C}}(x,T_0,P_0)$ of the classical system is available, the TI scheme can be performed to obtain the Gibbs free energy $G_{\mathcal{A}}(x,T_0,P_0)$ at *ab initio* level [22,37]. We note the volumes of $\mathcal{A}$ and $\mathcal{C}$ systems as $V_{\mathcal{A}}$ and $V_{\mathcal{C}}$ at the pressure $P_0$. The Gibbs free energy can be written as follows,

$$G_{\mathcal{A}}(x,T_0,P_0) - G_{\mathcal{C}}(x,T_0,P_0) = F_{\mathcal{A}}(x,T_0,V_{\mathcal{A}}) + P_0V_{\mathcal{A}}(x,T_0,P_0) - F_{\mathcal{C}}(x,T_0,V_{\mathcal{C}}) - P_0V_{\mathcal{C}}(x,T_0,P_0), \quad \text{(A8)}$$

where $F_{\mathcal{A}}$ and $F_{\mathcal{C}}$ are the Helmholtz free energy of $\mathcal{A}$ and $\mathcal{C}$ systems, respectively. Here, $F_{\mathcal{A}}(x, T_0, V_{\mathcal{A}}) - F_{\mathcal{C}}(x, T_0, V_{\mathcal{C}})$ term can be written as

$$F_{\mathcal{A}}(x, T_0, V_{\mathcal{A}}) - F_{\mathcal{C}}(x, T_0, V_{\mathcal{C}}) = \big(F_{\mathcal{A}}(x, T_0, V_{\mathcal{A}}) - F_{\mathcal{C}}(x, T_0, V_{\mathcal{A}})\big) + \big(F_{\mathcal{C}}(x, T_0, V_{\mathcal{A}}) - F_{\mathcal{C}}(x, T_0, V_{\mathcal{C}})\big). \tag{A9}$$

Because $P = -\left(\frac{\partial F}{\partial V}\right)_T$, we can write

$$F_{\mathcal{C}}(x, T_0, V_{\mathcal{A}}) - F_{\mathcal{C}}(x, T_0, V_{\mathcal{C}}) = -\int_{V_{\mathcal{C}}}^{V_{\mathcal{A}}} P_{\mathcal{C}}(x, V, T_0) dV \quad , \tag{A10}$$

We define $f_{PV}$ as

$$f_{PV}(x, T_0, P_0) = P_0 V_{\mathcal{A}} - P_0 V_{\mathcal{C}} - \int_{V_{\mathcal{C}}}^{V_{\mathcal{A}}} P_{\mathcal{C}}(x, V, T_0) dV \quad , \tag{A11}$$

where $P_{\mathcal{C}}(x, V, T_0)$ is the equation of states of the solution for system $\mathcal{C}$. The $f_{PV}$ term requires the equilibrium volumes of the solution for systems $\mathcal{A}$ and $\mathcal{C}$, respectively. We also define $F_{\mathcal{A}}(x, T_0, V_{\mathcal{A}}) - F_{\mathcal{C}}(x, T_0, V_{\mathcal{A}})$ as the $f_{TI}$ term, which can be calculated by TI using the classical system as the reference state [22], i.e.

$$f_{TI}(x, T_0, P_0) = F_{\mathcal{A}}(x, T_0, V_{\mathcal{A}}) - F_{\mathcal{C}}(x, T_0, V_{\mathcal{A}}) = \int_0^1 < U_{\mathcal{A}}(x, T_0, V_{\mathcal{A}}) - U_{\mathcal{C}}(x, T_0, V_{\mathcal{A}}) >_{\lambda, NVT} d\lambda, \tag{A12}$$

where $U_{\mathcal{A}}$ and $U_{\mathcal{C}}$ are the internal energy of solutions for systems $\mathcal{A}$ and $\mathcal{C}$, respectively. $\langle \cdot \rangle_{\lambda, NVT}$ is the ensemble average of internal energy over configurations sampled in the canonical ensemble with the force field $U = (1 - \lambda)U_{\mathcal{C}} + \lambda U_{\mathcal{A}}$. The subscript *NVT* indicates the constant conditions of volume ($V_{\mathcal{A}}$) and temperature ($T_0$) in the MD simulations of TI.

Combining Eqn. (A8)-(A12), the *ab initio* Gibbs free energy for liquid and superionic solutions can be obtained as

$$G_{\mathcal{A}}(x, T_0, P_0) = G_{\mathcal{C}}(x, T_0, P_0) + f_{PV}(x, T_0, P_0) + f_{TI}(x, T_0, P_0), \tag{A13}$$

With Eqn. (A13), $G_{\mathcal{A}}$ can be calculated for liquid, superionic *hcp*, and *bcc* at any oxygen composition, temperature, and pressure.

*Simulation details*—Classical MD simulations were performed using the Large-scale Atomic/Molecular Massively Parallel Simulator (LAMMPS) code [48]. The embedded-atom method (EAM) potential was developed to simulate superionic Fe-O systems under Earth's core conditions. The time step in the classical MD simulation was 1.0 *fs.* The Nosé-Hoover thermostat and barostat obeying modular invariance [49] were applied with the damping time $\tau = 0.1\, ps$. 323 GPa was chosen as the target pressure to maintain consistency with previous free energy calculations [22,43,46,50].

*Ab initio* molecular dynamics (AIMD) simulations were performed using the Vienna *ab initio* simulation package (VASP) [51,52]. The projected augmented wave (PAW) method [53] was used to describe the electron-ion interaction. The generalized gradient approximation (GGA) in the Perdew-Burke-Ernzerhof (PBE) form [54] was employed for the exchange-correlation energy functional. Tests with other exchange–correlation functionals (PBEsol and SCAN) suggest that the PBE functional provides the best description of Fe's equation of state (EOS) at high pressures, compared to experimental data (Supplementary Note 7 [33]). The electronic entropy was included using the Mermin functional [55,56] and the electronic temperature in the Mermin functional was the same as the ionic temperature. Supercells with 250, 288~293 and 250~254 atoms were used for liquid, superionic *hcp* and superionic *bcc* phases, respectively. The size of the simulation cell is found to have a negligible effect on the free energy calculation of *bcc* Fe (Supplementary Fig. S10 [33]). The time step in AIMD and TI was 1 *fs*. PAW potential with $3d^7 4s^1$ valence electrons (noted as PAW8) was used for Fe in the AIMD and TI. PAW potential with $3s^2 3p^6 3d^7 4s^1$ valence electrons (noted as PAW16) was used in the FEP. PAW potential with $2s^2 2p^4$ valence electrons was used for O. The plane-wave cutoff was 400 eV for PAW8-Fe and 750 eV for PAW16-Fe. The $\Gamma$ point was used in the AIMD. A dense Monkhorst-Pack ***k***-point mesh of $2 \times 2 \times 2$ was adopted for superionic and liquid phases to achieve a high DFT accuracy in the FEP calculations. The PAW16 potential produces an EOS consistent with all-electron calculations for both *hcp* and *bcc* Fe. (Supplementary Note 7 [33]). For a target pressure, the lattice parameters of *bcc*, *hcp*, and liquid phases were adjusted for each temperature and composition to ensure the pressure fluctuated around the target value by less than 0.5 GPa within 5 *ps* of simulations. The enthalpy data were collected from AIMD lasting more than 10 *ps*.